\documentclass[12pt,preprint]{aastex}
\usepackage{amsmath}
\shorttitle{Spherical Panoramas}
\shortauthors{Kent}

\begin{document}

\title{Spherical Panoramas for Astrophysical Data Visualization}

\author{Brian R. Kent}
\affil{National Radio Astronomy Observatory\altaffilmark{1}\\ 520 Edgemont Road, Charlottesville, VA, 22903, USA\\ bkent@nrao.edu}

\altaffiltext{1}{The National Radio Astronomy Observatory is a facility of the  
National Science Foundation operated under cooperative agreement by  
Associated Universities, Inc. }

\begin{abstract}

Data immersion has advantages in astrophysical visualization.  Complex
multi-dimensional data and phase spaces can be explored in a seamless
and interactive viewing environment.  Putting the user in the data
is a first step toward immersive data analysis.
We present a technique for creating 360$^{\circ}$ spherical panoramas with
astrophysical data.  The three-dimensional software package Blender and the Google Spatial
Media module are used together to immerse users in data exploration.
Several examples employing these methods exhibit how the technique works
using different types of astronomical data.\\

\textbf{Accepted for publication in the journal PASP Special Focus Issue \textit{Techniques and Methods for Astrophysical Data Visualization.}}

\end{abstract}


\section{Introduction}
\label{intro}

Data presentation in astronomy has expanded beyond the traditional 
two-dimensional plots to include multi-variable analysis, immersive three-dimensional (3D) models 
and exploratory analysis on a multitude of hardware platforms.  
Many forms of astronomical data can benefit from alternative forms of data exploration.  Data cubes can be rotated in 3D, fly-throughs of planetary landscapes can be explored, and 
all-sky maps can be examined.
Exploratory 3D presentations give users a different perspective on their data.
Traditional desktop and laptop configurations as well as hand-held mobile devices
can be used as the hardware mechanism for these data explorations.
The techniques can be employed in both research-level visualization and can deliver easy access to materials for education and public outreach \citep{2013PASP..125..731K}.

Panoramic imaging in the physical domain is accomplished with a tripod mounted
camera to encompass a larger field of view (FOV) than is visible through the lens.  Multiple images are obtained while
attempting to minimize parallax errors between images \citep{Derrien2000}.  In some cases, four to 
six ultra wide-field cameras
are used to record video in every possible direction.  Software can then merge these images and video encompassing a
FOV into a cylindrical or spherical panorama \citep{Ozawa2012}.  Computer hardware and screen resolution
along with network bandwidth are now sufficient to allow 360$^{\circ}$ spherical panoramic video: a constant
stream of panoramic images obtained with multiple wide-field cameras that gives a user
control over the view while the video advances.  Natural movement to view the scene can be accomplished
with accelerometer-enabled hand-held devices or with desktop browsers.

We present a methodology overview for creating still and video spherical panoramas using astronomical data.  Section~\ref{panorender} gives the reader a brief overview of rendering
with Blender and the Google Spatial Media module.  In section~\ref{methodssection} we describe the math and techniques behind building a spherical panorama.  Section~\ref{examples} outlines examples utilizing the libraries and techniques described in this paper.  Section~\ref{summary} summarizes the techniques.

\section{Panoramic Rendering}
\label{panorender}

Traditional data presentation revolves around graphs, charts, tables, and other forms of two-dimensional media.  They have a critical role in research publications.
Complex or multi-dimensional data can benefit from a more advanced data display mechanism.  The ability to immerse users in their data has a number of technical challenges that can be accommodated with recent advances in both hardware and software.  Data can be presented within a 3D space, including data cube volume rendering, planetary surface maps, and astronomical catalogs \citep{2011PASA...28..150H, 2013PASP..125..731K}.  This paper introduces astronomers to spherical panoramas as a data display mechanism.  It is advantageous to use immersive spherical panoramas as they allow easy and natural control of the viewing space, intuitive and minimalist user interfaces, and cross-platform support across a range of hardware and display software.

Projection algorithms in astronomical FITS images have been well curated and documented \citep{2001A&A...376..359H}.  The growing size of CCD cameras required new drizzling algorithms and reprojection models that paved the way for mosaicking and fast, large, wide-field projection and coordinate transformation algorithms \citep{2002PASP..114..144F, 2004PASP..116..971M}.  Montage is a recent example of image mosaicking and projection code that has been very successful on this front \citep{2007HiA....14..621B, Jacob2009}.  While this paper focuses on adding value to existing images and data, it is important to acknowledge the algorithms that go into making the products from large projects.  Astronomical surveys produce wide-field imaging science-ready data products and catalogs \citep{2009ApJS..182..543A, 2006AJ....131.1163S, 2008AJ....136..713K}.  Wide-field radio surveys require special algorithms where mosaicking is done in the interferometric \textit{uv}--plane \citep{2012CRPhy..13...28T, 2014MNRAS.444..606O} as opposed to the imaging plane.

With wide-field mosaics across multi-wavelength regimes increasing in size, resolution, sensitivity and detail, it is useful to create immersive spherical panoramas where the user is \textit{inside} the data. Panoramic photography and imaging play an important role in interactive user interfaces, data analysis, and 3D reconstruction \citep{Peleg2001, Gledhill2003}.  More recently, stereoscopic vision, virtual reality, and augmented reality applications can combine to supplement data viewing in geographic and mapping software \citep{Azuma1997, Azuma2001, Azuma2004, Wither2011}.  This has applications in haptic feedback devices, 3D caves, and other forms of virtual reality \citep{Sanchez2005}.  Displaying 3D data on traditional two-dimensional (2D) displays (a computer or a tablet) requires certain considerations when projecting images~\citep{Gilbert2012}.  Further applications include stereo vision in 3D environments with headgear apparatus~\citep{Gasparini2013}.  We focus here on the simplest use of easy to obtain hardware: desktops, laptops, and mobile devices.  Used in conjunction, the software package Blender combined with the Google Spatial Media Library make a powerful combination for a researcher to animate and display their data for other users.

\begin{figure}
\epsscale{0.98}
\plotone{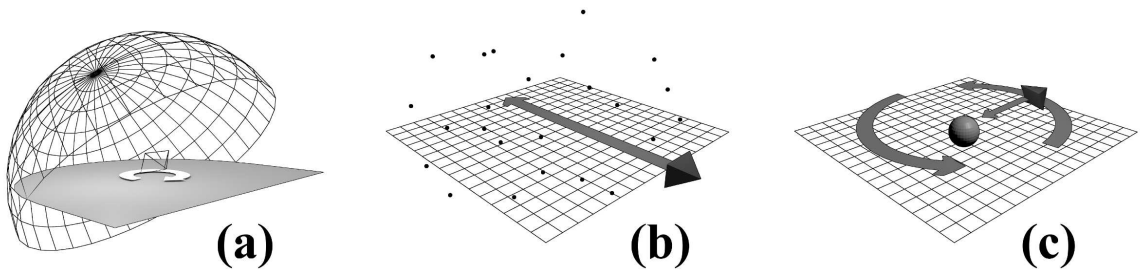}
\caption{The three spherical panorama scenarios described in this paper.  
(a) A static view inside
the celestial sphere where a user can move their browser or mobile device over all 4$\pi$ steradians.
(b) A straight path of a camera flying through a 3D catalog of astronomical objects.
(c) The camera revolving around a central object.  The camera tracks the object with its virtual focal
plane always perpendicular to the object normal vector.
\label{blender_paper_geometry_diagram}}
\end{figure}

We outline spherical panoramas described in this paper that can fit into one
of three scenarios~(Figure~\ref{blender_paper_geometry_diagram}).  
An example from each of these scenarios will be described in section~\ref{examples}.

\begin{itemize}
\item \textit{Static spherical panoramas}.  This type is best exemplified by
a user stationary in the middle of the celestial sphere.  All-sky maps work
well with this scenario.  It requires a single frame and can be executed with short render times.
\item \textit{Single pass fly-throughs}.  Imagine this scenario as moving through
 a 3D dataset on a predetermined track.  As the user moves along the track at a fixed speed,
 they are free to look about in any direction.
\item \textit{Orbit flyovers}.  The scenario is geared toward data where one item is in focus, such as a particular area or object in a simulation or map feature.
\end{itemize}

\subsection{Blender}
\label{blender}

Blender supports a wide range of 3D visualization and animation scenarios\footnote{\url{https://www.blender.org/}}.  
It is maintained by the Blender Foundation as open-source software.  The software has been adapted for use in a number of astronomy applications with easily accessible primers and tutorials \citep{2013PASP..125..731K, 2015svb..book.....K,  2015A&C....13...67T, 2016A&C....15...50N, Garate2017}.  The novice user is referred to those works and the references therein when using Blender
for the first time with their data.  Blender versions 2.64a and later will work with the examples
and techniques outlined in this paper.

\begin{figure}
\epsscale{0.95}
\plotone{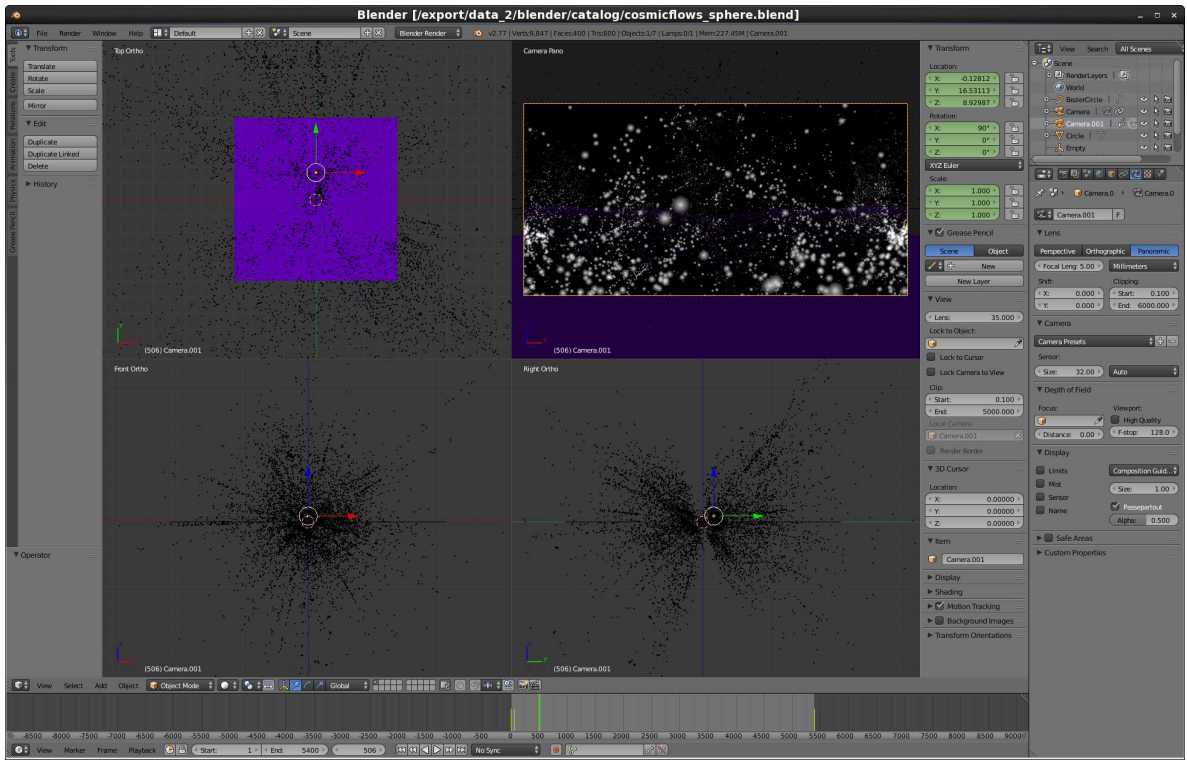}
\caption{Blender graphical user interface (GUI) setup for panoramic rendering.  This particular view is using a 3D catalog of galaxies from \citet{2009AJ....138..323T}.  The display gives the user a view along all three display axes.  When rendered
with panoramic settings, the result of a 360$^{\circ}\times$180$^{\circ}$ FOV is shown
in the upper right panel.  Reproduced from the Blender GUI under the GNU General Public License (GPL) (https://www.blender.org/about/). \label{blendergui}}
\end{figure}

Three-dimensional graphic suites like Blender require a workflow
of data selection, modeling, and animation.  Other software packages referred to in \citet{2013PASP..125..731K} including Lightwave 3D, Maya, and 3D Studio Max are popular in the graphics industry.  While the internal working models and graphical user interfaces may differ between packages, Blender supports data import from all of those packages.  As with any data reduction or graphical display package and depending the complexity of data, it is best
to begin learning the package with a simple dataset such that the developed data model can be built upon for reuse.
The user's proficiency can quickly improve with the tutorials and examples provided at the end of this paper.  To aid
in the ease of scripting, the Python application program interface (API) allows interfacing with popular scientific astrophysics libraries \citep{2013A&A...558A..33A} such that a variety
of astronomical data (ASCII catalogs, FITS images, etc.) can be brought into the environment.  We choose to use Blender because its built-in engine capabilities allow for fast panoramic rendering of a 3D scene.  In addition, the new \textit{Cycles} rendering engine allows for a node-based compositor control of many of Blender's graphical user interfaces (GUIs) and features~(Table~\ref{enginetable}).
The camera controls in Blender give a 360$^{\circ}\times$180$^{\circ}$ FOV
that can be used as a still image in a static spherical panorama~(Figure~\ref{blendergui}).

\subsection{Google Spatial Media}
\label{googlemedia}

The Google Spatial Media module is a collection of specifications and tools for spherical 360$^{\circ}$~video and spatial audio. The module can take rendered output from Blender and generate spherical panorama video.  This is accomplished with a metadata injector that adds information on the file type and FOV to the header.  Audio can also be included in the metadata, but it will not be discussed in this work.  The module source code is available from a GitHub repository\footnote{\url{https://github.com/google/spatial-media}}.  The module is available as a stand-alone executable for Windows and Mac OS or as a Python module for Linux/Windows/Mac OS.  By adding to the video metadata header, the output can then be exported to YouTube\footnote{\url{http://www.youtube.com}}.
Within YouTube the player becomes aware of the data content and gives the user a set of controls with which to navigate the environment.  Users can use a mouse to click and drag or move their tablet or mobile device naturally to view the scene.
The preprocessing by Blender to create a full spherical view covering 4$\pi$ steradians is critical before running it through
this module.  It is assumed that the metadata injection carried out by the module is being applied to to a full spherical panoramic view.

The social media website Facebook also has the capability of supporting 360$^{\circ}$ video,
but the metadata injection must be inserted by the user if the video has not been
already created by a spherical camera system.  This creates a complication that
is solved for the user by using the Google Spatial Media module with the added
benefit of the Python interface.   Other 3D VR/AR (virtual reality/augmented reality)
applications such as Oculus Rift or Vive require rather steep hardware requirements.
The scheme presented in this paper is aimed at a researcher utilizing 
hardware already available as well as open-source software for the processing.

\subsection{Hardware and Software Requirements}
\label{hardwaresubsection}

\begin{figure}
\epsscale{0.9}
\plotone{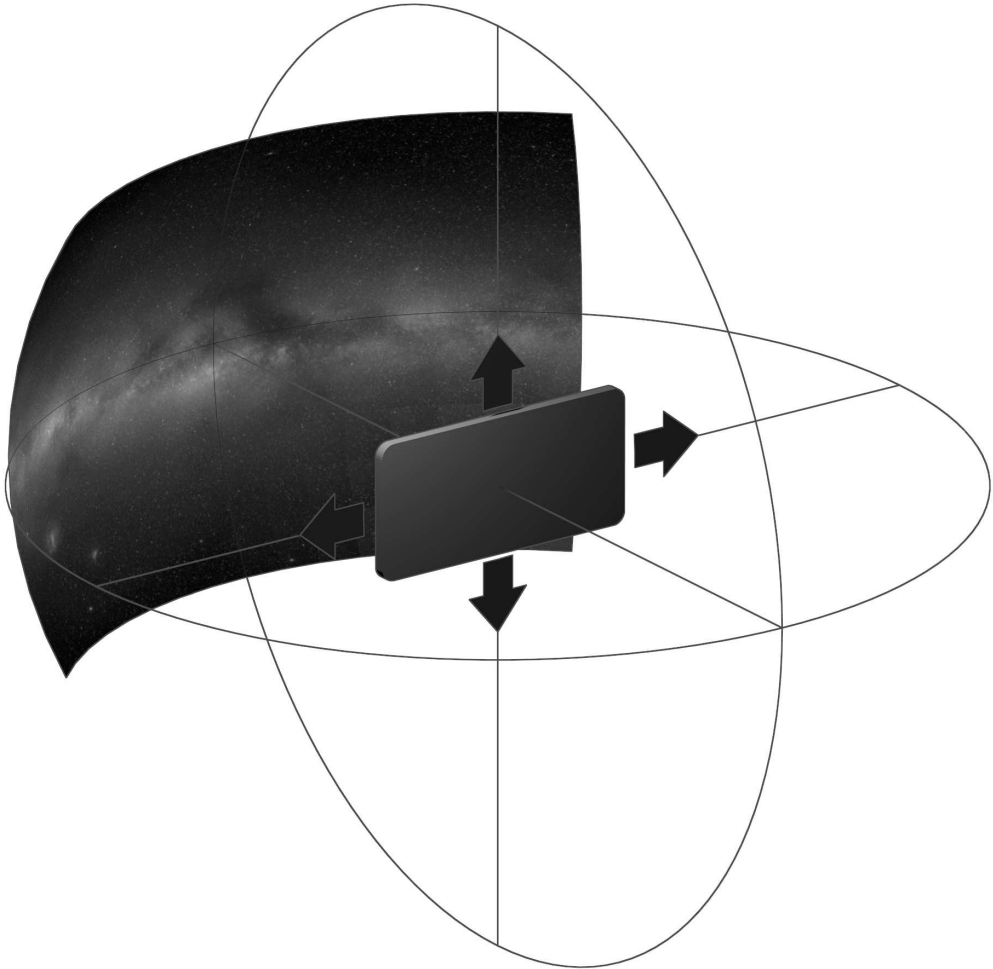}
\caption{Mobile device or tablet can be rotated in 3D to view any direction in the 3D space.  In addition, during spherical panoramic videos, the camera can be translated in the \textit{x}, \textit{y}, or \textit{z} directions along a predetermined path in Blender.  The astronomical image subtending the sphere shows the device's FOV.\label{phone}}
\end{figure}

The Google software interface coupled with Blender and YouTube allows for an easy implementation of a spherical panorama video on a variety of hardware platforms.  The results can be viewed with a mouse-desktop/laptop setup with the Chrome\footnote{\url{http://chrome.google.com}} or Firefox browser\footnote{\url{http://http://www.mozilla.org/firefox}}.  Tablets and mobile devices with three-axis accelerometers are where this technique's forte really lies.  Using this technique on such devices allows for an immersive data discovery session.  A user can move within the dataset as the video plays or be located in a static position~(Figure~\ref{phone}).  The device itself can be moved or the user can navigate with their hands.  If the camera is moved during a spherical panorama video, users can track a particular object by pointing their device in different directions.  User device rotation is the controllable element in this scheme.  No translation of the mobile device will have an effect on the viewer's perspective.  The technique works equally well on both iOS and Android mobile operating systems.

\section{Methods}
\label{methodssection}

Images can be combined and projected from an equirectangular, cylindrical equal area or Hammer-Aitoff projection.  
The first two may be necessary
in certain cases where data is not available to encompass the entire sphere of visibility.
An equirectangular projection conversion is simply defined with Cartesian coordinates \textit{x,y}, and angles $\phi,\theta$ as
\begin{equation}
x = \phi
\end{equation}
\begin{equation}
y = \theta
\end{equation}

\begin{figure}
\epsscale{1.0}
\plotone{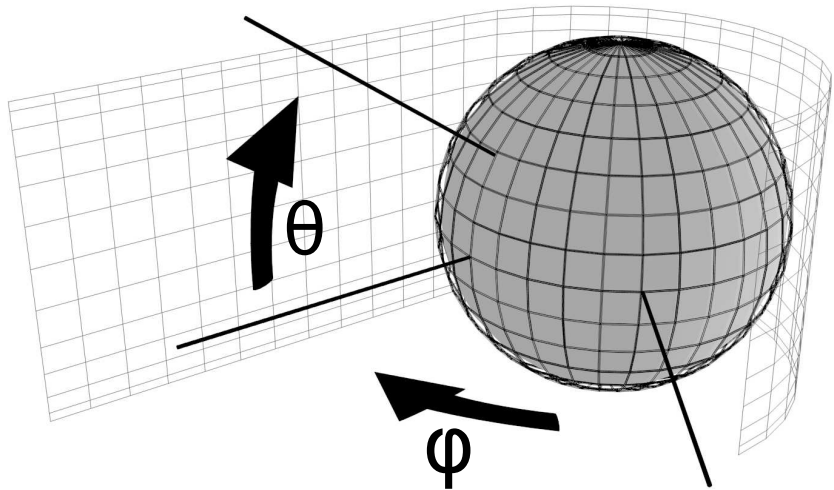}
\caption{Wrapping of a Lambert equal area projection on a sphere with polar angle $\theta$ and azimuthal angle $\phi$.\label{cylinder_geometry}}
\end{figure}

A cylindrical equal area projection can be defined as
\begin{equation}
x = \phi
\end{equation}
\begin{equation}
y = \frac{180^{\circ}}{\pi} \frac{\rm{sin~}\theta}{\lambda}
\end{equation}
where $\phi$ is the azimuthal angle and $\lambda$ is the radius of the projected cylinder.  We will
assume a scaling of $\lambda = 1$ in this work's examples for simplicity (known as a Lambert's equal area projection).
A simple cylindrical panorama setup with a Lambert's equal area projection is shown in Figure~\ref{cylinder_geometry}.

For all-sky maps, image reprojection is the most computationally intensive part of image processing. Tools such as Montage assist with this \citep{Makovoz2004, Jacob2009}.  All-sky maps are readily available from a large number of astronomical surveys.  The Hammer-Aitoff to spherical
formalism is given as
\begin{equation}
\label{sphere_eq1}
x = 2\gamma \rm{~cos~}\theta \rm{~sin} \frac{\phi}{2}
\end{equation}
\begin{equation}
\label{sphere_eq2}
y = \gamma \rm{~sin~}\theta
\end{equation}
where
\begin{equation}
\label{sphere_eq3}
\gamma = \frac{180^{\circ}}{\pi}\sqrt{\frac{2}{1+\rm{cos~}\theta \rm{~cos}(\phi/2)}}
\end{equation}

These maps and projection definitions are well defined by \citet{2002A&A...395.1077C}.
These transformations will be employed in section~\ref{examples}.

Once a FITS image map is generated, it can be imported into Blender by way of AstroPy and its associated FITS manipulation modules.  Planetary surface maps and 3D catalogs can be brought into the Blender environment as described in \citet{2013PASP..125..731K, 2015svb..book.....K}.  Here we focus on the methods and geometry of virtually filming these data for spherical panoramas.

\begin{figure}
\epsscale{1.0}
\plotone{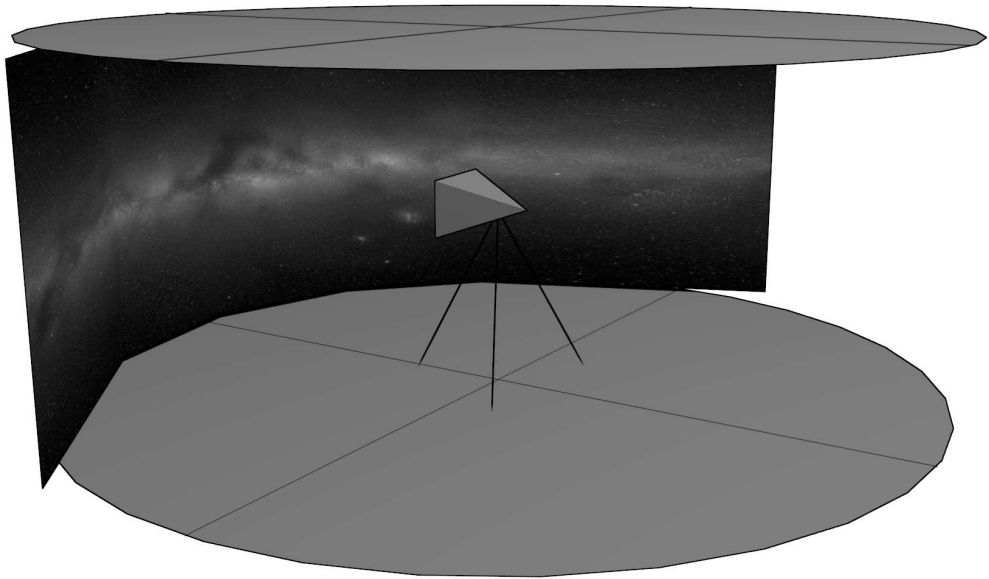}
\caption{Diagram of a schematic cylindrical panorama setup.\label{panosetup_cylinder}}
\end{figure}

\begin{figure}
\epsscale{0.7}
\plotone{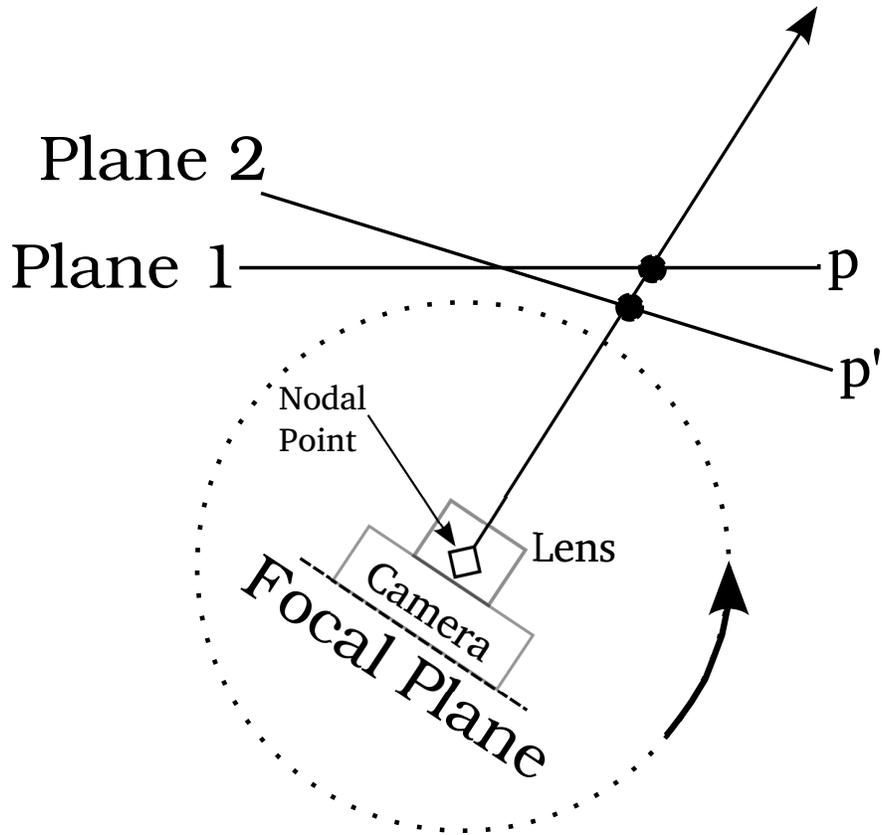}
\caption{Diamond symbol indicates the virtual nodal point in Blender at which the camera rotates to photograph the panorama.
The nodal point and focal plane show the relative locations of an actual physical camera setup.
In this particular configuration, two planes are overlapping and are viewed with the virtual camera.
\label{nodaldiagram}}
\end{figure}

We can consider the method of virtually photographing our \textit{scene} with data in a similar fashion to the way a camera is set up to image in a panoramic mode (Figure~\ref{panosetup_cylinder}).  While the internal Blender mechanics automate
this process with the built-in camera object, it is useful to understand some of the underlying mathematics.   The projective mapping between two planes is given by
\begin{equation}
\begin{bmatrix}
         wx^{\prime} \\
         wy^{\prime} \\
         w 
        \end{bmatrix}
=
\begin{bmatrix}
         h_{11} & h_{12} & h_{13} \\
         h_{21} & h_{22} & h_{23} \\
         h_{31} & h_{32} & h_{33} 
        \end{bmatrix}
\begin{bmatrix}
         x \\
         y \\
         1 
        \end{bmatrix}
\end{equation}
which is summarized as
\begin{equation}
\rm{p}^{\prime} = \rm{H~} p
\end{equation}
where \textbf{H} is the \textit{homography matrix} that transforms \textbf{p} to our plane \textbf{p$^{\prime}$}.  The formalism for panoramic homography is reviewed in \citet{Hartley2004}.
When projecting images for viewing, it is important to understand that we are not rotating
the FOV about the focal plane, but around the \textit{nodal point} for the 
exit pupil that sits in front of the plane (Figure~\ref{nodaldiagram}).  The nodal point for a physical camera is the correct pivot point in panoramic imaging and allows us to minimize parallax errors ~\citep{ZhangLiu2014}.
A virtual camera in a 3D scene can be thought of as a point from which rays originate constrained by a FOV.
The rays in a virtual camera's 360$^{\circ}$ FOV are unconstrained (see \citet{Glassner1989} for a comprehensive review on ray tracing).  Stereo imaging is not within the scope
of this paper, but it will utilize two origin points for the rays separated by an inter-pupillary distance, also refereed to as a stereo baseline \citep{Peleg2001}.

With this information in mind, the broad workflow for creating our examples will work as follows.  Each example will give details
specific to that given scenario and will utilize the following seven steps.  The seventh stage will be omitted from
the examples as it is uniform to all three, where we use the Google Spatial Media module.

\begin{figure}
\epsscale{0.9}
\plotone{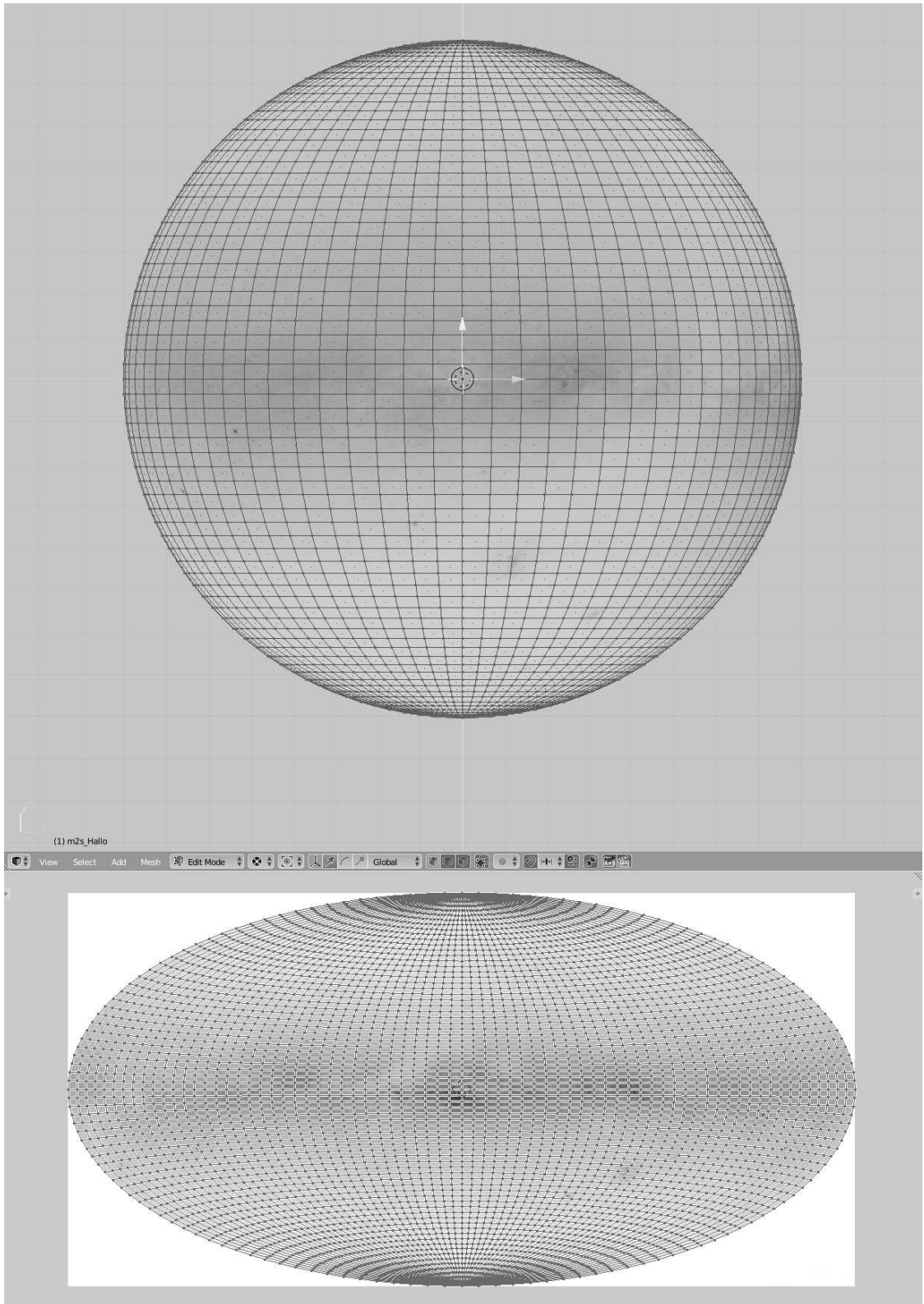}
\caption{Diagram from the Blender interface shows the \textit{uv}-wrapping of a Hammer-Aitoff all-sky map onto the celestial sphere when creating a spherical panorama.  Reproduced from the Blender GUI under the GNU General Public License (GPL) (https://www.blender.org/about/). \label{uvcoverage}}
\end{figure}

\begin{enumerate}

\item  \textit{\textbf{Import data}}.  Maps, textures, and catalogs are brought into the environment.  Scripts for these processes are available in \citet{2013PASP..125..731K}.
\item  \textit{\textbf{Set uv-mapping}}.  For astronomical images and maps that need to be projected into the 3D environment, \textit{uv}-unwrapping can be utilized~(Figure~\ref{uvcoverage}) using
equations~\ref{sphere_eq1}--\ref{sphere_eq2}.
\item  \textit{\textbf{Configure data scene}}. The data scene configuration includes any illumination, texture, and material objects that are needed in the visualization.
\item  \textit{\textbf{Set camera path}}.  The path is configured with a B{\'e}zier curve if required.  The camera can be moved freely as determined by the user or restricted
to initially focus on a particular area or object of the data.
\item  \textit{\textbf{Configure camera}}.  The camera settings can be configured to maximize the FOV depending
on how much coverage is yielded by the data (for instance, only a partial set of data in an all-sky map).  This requires a panoramic camera selection in either the default rendering engine or the new Cycles engine.
\item  \textit{\textbf{Render Scene}}.  The scene should be rendered at 4K or higher since the spherical panorama image must cover 4$\pi$ steradians \citep{Silva2016}.  The best output resolution should be used at 3840~$\times$~2160 pixels.
\item  \textit{\textbf{Configure metadata}}.  The rendered video can be imported into the Google Spatial Media module as a stand-alone binary or as part of a Python script.

\end{enumerate}

%



\section{Examples}
\label{examples}

The examples described in this section follow data import and manipulation procedures
outlined in \citet{2013PASP..125..731K, 2015svb..book.....K}.
All data manipulation examples start with the default file that opens with Blender
upon startup; no changes need to be made to the default configuration.
Most workflows in these examples require a number of mouse/keyboard binding combinations.\footnote{References for keyboard entry and model manipulation can be found at~\url{http://www.cv.nrao.edu/$\sim$bkent/blender/tutorials.html}}

The first example uses a static view inside a celestial sphere.  The second and third examples
both utilize a moving camera with two different data sets, an astronomical catalog where the camera flies through with a linear path
and a 3D planetary surface with a circular path where the camera can focus on a particular geographic feature.

\subsection{All-sky maps}

\begin{figure}
\epsscale{0.9}
\plotone{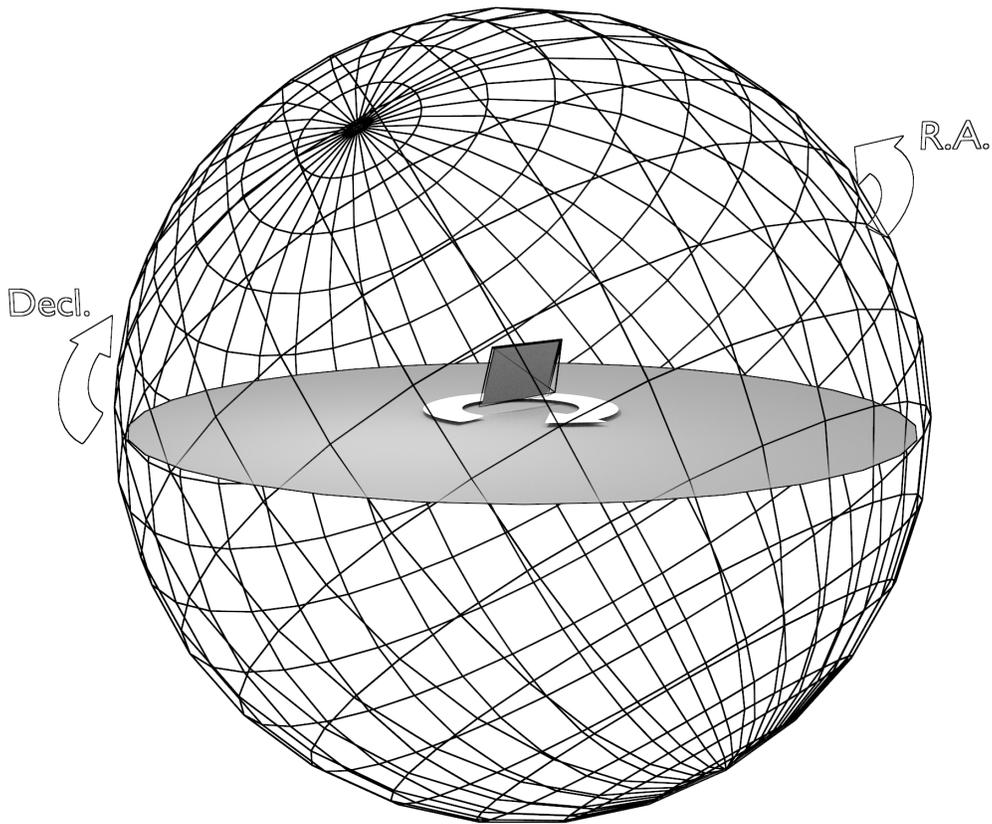}
\caption{Spherical geometry setup of a virtual camera inside the celestial sphere, showing R.A. and declination.  All-sky maps are projected onto the sphere from a Cartesian or Hammer-Aitoff map and filmed into a spherical panorama by the camera.\label{spherical}}
\end{figure}

All-sky map data lend themselves very well to the spherical panorama paradigm.  While any astronomical FITS
image can be used when properly placed on the celestial sphere, wider sky coverage produces a more pronounced effect.
All-sky maps cover a complete 4$\pi$ steradians FOV of spatially well-sampled data
over a wide range of wavelengths.  For this example, we use a static camera view where the user does not translate.  
The user is essentially placed inside the center of a celestial sphere and can rotate to view everything around them (Figure~\ref{spherical}).
Data from \textit{Skyview} can be used, obtaining images centered on the galactic center ($l=0^{\circ}, b=0^{\circ}$) in a Cartesian or
Hammer-Aitoff projection \citep{1994ASPC...61...34M, 2009PASP..121.1180M}.  Oversampling the image in a 2:1 or 16:9 aspect ratio with a 360$^{\circ}\times$180$^{\circ}$ FOV is best; the end product will be viewable on different resolutions with various hardware.

The workflow for this example is as follows:

\begin{enumerate}

\item  Importing data can be accomplished with \textbf{astropy.io.fits.open} or with a high-resolution JPEG, PNG, or TIFF image.  Images can be added via the UV/Image Editor (see the lower panel of Figure~\ref{uvcoverage}).  The data scene requires that a \textit{uv} sphere mesh be added to the scene by pressing the key combination \textbf{SHIFT-A}.

\item  Pressing the \textbf{TAB} key puts the main viewer in \textit{Edit} mode.  

The \textbf{A} key followed by the \textbf{U} key is used to map the image onto the mesh model.

\item  A material and texture can be added to the mesh sphere with an Emission shading value of 1.0, Image set to the 
loaded map and Coordinates set to \textbf{UV}.

\item  The camera will be static in this example, rotated to ($x_{\theta}, y_{\theta}, z_{\theta}$) = ($0^{\circ}, 0^{\circ}, 0^{\circ}$) facing the galactic center at ($l=0^{\circ}, b=0^{\circ}$).

\item  The Cycles engine can be used to our advantage here, setting the FOV to 180$^{\circ}$.  

\item  The resolution output should be set to 3840~$\times$~2160 pixels, the highest resolution currently available on YouTube\footnote{https://support.google.com/youtube/answer/6375112}.  
The video output should be set to the MPEG with MPEG-4 encoding and a bitrate of 8000 kilobits per second.

\end{enumerate}

\subsection{Catalog fly-through}

In this example, we tour a galaxy catalog \citep{2009AJ....138..323T} where the
user is on a predefined moving track through 3D space.  
This automated constant speed translation is coupled with the 
user controlled rotation as the video visualization plays.
Translating schemes like this are good for scenarios in which various aspects of the data
need to be given a closer look. It also may be the case that the dataset covers a large parameter
space, thus traversing a particular axis is advantageous.

\begin{figure}
\epsscale{0.9}
\plotone{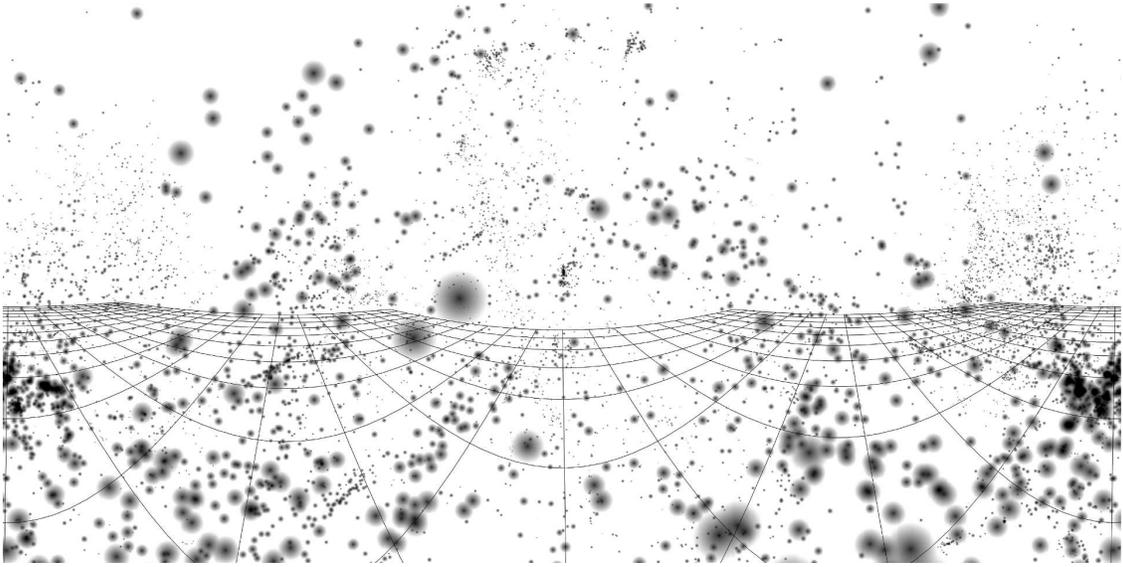}
\caption{360$^{\circ}\times$180$^{\circ}$ image showing the output from Blender with its built-in panoramic camera
The data shown are a 3D catalog of galaxies from \citet{2009AJ....138..323T}. The resulting image will be processed by the Google Spatial Media module.\label{catalog_pan}}
\end{figure}

\begin{enumerate}

\item  Data can be imported from ASCII text files, XML or binary tables using standard Python modules.

\item  3D catalogs are best rendered as a single mesh with disconnected vertices in \textit{Edit} mode.

\item  Halo \textbf{Materials} are best used with a Emission shading value of 1.0 to create point sources for each catalog object.

\item  A linear path is set up such that the camera will traverse 75\% of the full width at half maximum (FWHM) of the width of the catalog's radial distribution.  This allows
reasonably uniform coverage of the viewing sphere as the user moves through the 3D catalog space.

\item  Single \textbf{Halo} materials are not supported in the Cycles engine; we use the standard Render engine provided with Blender.

\item  The resolution output should be set to 3840~$\times$~2160 pixels (Figure~\ref{catalog_pan}).  The video output should be set to the MPEG with MPEG-4 encoding and a bitrate of 6000 to 8000 kilobits per second,
depending on the length of the translation.  A trade off can be made between a longer video
and a slightly smaller bitrate.

\end{enumerate}

\subsection{Planetary Terrain Surface}

\begin{figure}
\epsscale{0.9}
\plotone{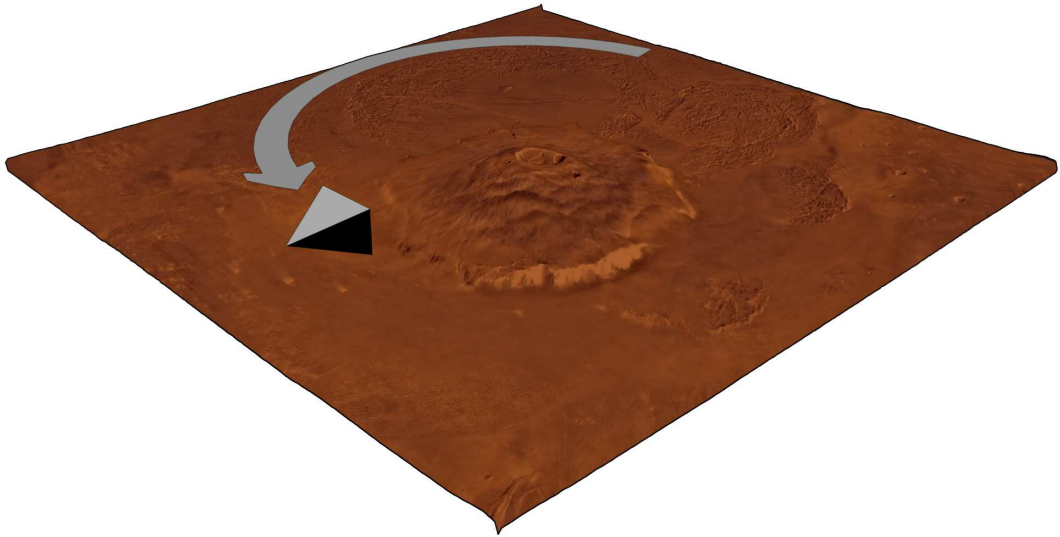}
\caption{Camera and its circular orbit path are shown in a simulated Martian 
terrain flyover.  The camera default will point to Olympus Mons in this scenario. \label{terrain}}
\end{figure}

With this scenario, we can render a panoramic view of the Martian shield volcano Olympus Mons.  The camera path will encircle the geographic feature, allowing the user to pan the camera in any direction during the animation.  The data discussed in \citet{2015svb..book.....K} and \citet{Smith1999} can be used as an example, with layers used to build the terrain maps\footnote{\url{http://astrogeology.usgs.gov/tools/map}}~(Figure~\ref{terrain}).  As the camera revolves around the volcano,
the altitude is changed along a predetermined track of revolution.  Any kind of solid 3D model viewed
externally is well suited to this visualization scenario.


\begin{enumerate}

\item  Loading planetary surface images requires both satellite imaging as well as some 
form of altitude ranging.  These can be layered as a displacement modifier and texture for \textit{uv}-mapping.

\item  Surface data can be rendered as as a subdivided surface plane in \textbf{Edit} mode with a deformation modifier.

\item  A regular surface mesh material with an Emission shading value of 1.0 works best in this example.

\item  A B{\'e}zier circle can be used as a path animation combined with an Empty mesh (null vector) that will allow the camera to point toward the surface during the video.
       The camera is attached to this animated path.  The time frame of the path animation can determine the length of the visualization.
       
\item  The user can choose between Cycles and Render; either will work well in the scenario.

\item  The resolution output should be set to 3840~$\times$~2160 pixels.  The video output should be set to the MPEG with MPEG-4 encoding and a bitrate of 6000 to 8000 kilobits per second.

\end{enumerate}

\section{Summary}
\label{summary}

We have introduced a method for creating 360$^{\circ}$ spherical panoramas with
astrophysical data.  The technique employs the 3D software package Blender
and the Google Spatial Media module.  The outlined principles of panoramic
projection allow us to use these tools in concert for immersive
data exploration.  By moving within a full spherical view, a user
can interact with astronomical images and 3D catalogs, models, and maps.
The interaction works equally well in a traditional browser environment
as well as physical movement with mobile devices.
We provide examples for three scenarios that a user could build upon:
a static celestial sphere, a linear fly-through, and a revolving orbit
focusing on a single area or object.

Demonstration videos as well as sample files, Python scripts, and basic tutorials of Blender principles
are available at \url{http://www.cv.nrao.edu/$\sim$bkent/blender}

We gratefully acknowledge the careful reading by Tom Bridgman and Kel Elkins in improving
the content and clarity of this work for the astronomical community.


\begin{thebibliography}{99}

\bibitem[Abazajian et al.(2009)]{2009ApJS..182..543A} Abazajian, K.~N., Adelman-McCarthy, J.~K., Ag{\"u}eros, M.~A., et al.\ 2009, \apjs, 182, 543-558 

\bibitem[Astropy Collaboration et al.(2013)]{2013A&A...558A..33A} Astropy Collaboration, Robitaille, T.~P., Tollerud, E.~J., et al.\ 2013, \aap, 558, A33 

\bibitem[Azuma(1997)]{Azuma1997} Azuma, R.T. 1997, Presence: Teleoper. Virtual Environ. 6, 4

\bibitem[Azuma et al.(2001)]{Azuma2001} Azuma, R. T., Baillot, Y., Behringer, R., Feiner, S., Julier, S., \& MacIntyre, B. 2001, IEEE Comput. Graph. Appl. 21, 6

\bibitem[Azuma(2004)]{Azuma2004} Azuma, R. T, 2004,  ACM SIGGRAPH Course Notes, 26

\bibitem[Berriman et al.(2007)]{2007HiA....14..621B} Berriman, G.~B., Laity, A.~C., Good, J.~C., et al.\ 2007, Highlights of Astronomy, 14, 621 

\bibitem[Calabretta \& Greisen(2002)]{2002A&A...395.1077C} Calabretta, M.~R., \& Greisen, E.~W.\ 2002, \aap, 395, 1077 

\bibitem[Derrien \& Konolige(2000)]{Derrien2000} Derrien, S. and Konolige, K. 2000, Proc. of IEEE Workshop on Omnidirectional Vision, 85

\bibitem[Fruchter \& Hook(2002)]{2002PASP..114..144F} Fruchter, A.~S., \& Hook, R.~N.\ 2002, \pasp, 114, 144

\bibitem[G\'{a}rate(2017)]{Garate2017} G\'{a}rate, M. 2017, PASP, http://adsabs.harvard.edu/abs/2016arXiv161106965G

\bibitem[Gasparini \& Bertolino(2013)]{Gasparini2013} Gasparini, S. and Bertolino, P. 2013, IEEE Conference, Computer Vision and Pattern Recognition, doi: 10.1109

\bibitem[Gilbert et al.(2012)]{Gilbert2012} Gilbert, S. B., Booksuk, W., Kelly, J.W. 2012, Human Vision and Electric Imaging XVII, Proceedings of SPIE 8291: 1, doi:10.1117/12.912173.

\bibitem[Glassner(1989)]{Glassner1989} Glassner, A. S., 1989, An Introduction to Ray Tracing, Academic Press Ltd., London, UK, isbn: 0-12-286160-4

\bibitem[Gledhill et al.(2003)]{Gledhill2003} Gledhill, D., Tian, G. Y., Taylor, D., Clarke, D. 2003, Computers \& Graphics, 27, 3


\bibitem[Hanisch et al.(2001)]{2001A&A...376..359H} Hanisch, R.~J., Farris, A., Greisen, E.~W., et al.\ 2001, \aap, 376, 359 

\bibitem[Hartley \& Zisserman(2004)]{Hartley2004} Hartley, R.~I. and Zisserman, A. 2004, Multiple View Geometry in Computer Vision, Cambridge University Press, ISBN: 0521540518

\bibitem[{{Hassan} \& {Fluke}(2011)}]{2011PASA...28..150H}
{Hassan}, A., \& {Fluke}, C.~J. 2011, PASA, 28, 150

\bibitem[Jacob et al.(2009)]{Jacob2009} Jacob, Joseph C. and Katz, Daniel S. and Berriman, G. Bruce and Good, John C. and Laity, Anastasia C. and Deelman, Ewa and Kesselman, Carl and Singh, Gurmeet and Su, Mei, Hui and Prince, Thomas A. and Williams, Roy \ 2009, Int. J. Comput. Sci. Eng., 4, 73

\bibitem[Kent et al.(2008)]{2008AJ....136..713K} Kent, B.~R., Giovanelli, R., Haynes, M.~P., et al.\ 2008, \aj, 136, 713 


\bibitem[Kent(2013)]{2013PASP..125..731K} Kent, B.~R.\ 2013, \pasp, 125, 731 

\bibitem[Kent(2015)]{2015svb..book.....K} Kent, B.~R.\ 2015, 3D Scientific Visualization with Blender (IOP Publishing, Bristol, UK) Print ISBN: 978-1-6270-5611-3,    Online ISBN: 978-1-6270-5612-0


\bibitem[Makovoz(2004)]{2004PASP..116..971M} Makovoz, D.\ 2004, \pasp, 116, 971 

\bibitem[Makovoz \& Khan(2005)]{Makovoz2004} Makovoz, D. and Khan, I. 2005, Mosaicking with MOPEX, Proceedings of ADASS XIV, Vol. 347, 81

\bibitem[McGlynn \& Scollick(1994)]{1994ASPC...61...34M} McGlynn, T., \& Scollick, K.\ 1994, Astronomical Data Analysis Software and Systems III, 61, 34 

\bibitem[Mellinger(2009)]{2009PASP..121.1180M} Mellinger, A.\ 2009, \pasp, 121, 1180 


\bibitem[Naiman(2016)]{2016A&C....15...50N} Naiman, J.~P.\ 2016, Astronomy and Computing, 15, 50 


\bibitem[Offringa et al.(2014)]{2014MNRAS.444..606O} Offringa, A.~R., McKinley, B., Hurley-Walker, N., et al.\ 2014, \mnras, 444, 606 

\bibitem[Ozawa et al.(2012)]{Ozawa2012} Ozawa, T., Kitani, K.M., and Koike, H. 2012, Proc. of 3rd Augmented Human Internaional Conference, 20

\bibitem[Peleg et al.(2001)]{Peleg2001} Peleg, S., Ben-Ezra, M., Pritch, Y. 2001, IEEE Trans. on Pattern Analysis and Machine Intelligence 23, 3



\bibitem[Sanchez-Vives \& Slater(2005)]{Sanchez2005} Sanchez-Vives, M.V. \& Slater, M. 2005, Nature Rev. Neuroscience, 6, 4

\bibitem[Silva et al.(2016)]{Silva2016} Silva, R. M. A., Feij\'{o}, B., Gomes, P. B., Frensh, T., and Monteiro, D. 2016, ACM SIGGRAPH 2016 Proc, 70:1

\bibitem[Skrutskie et al.(2006)]{2006AJ....131.1163S} Skrutskie, M.~F., Cutri, R.~M., Stiening, R., et al.\ 2006, \aj, 131, 1163 

\bibitem[Smith et al.(1999)]{Smith1999} Smith, D. E. et al. 1999, American Association for the Advancement of Science, 284, 5419, doi: 10.1126/science.284.5419.1495

\bibitem[Tasse et al.(2012)]{2012CRPhy..13...28T} Tasse, C., van Diepen, G., van der Tol, S., et al.\ 2012, Comptes Rendus Physique, 13, 28 

\bibitem[Taylor(2015)]{2015A&C....13...67T} Taylor, R.\ 2015, Astronomy and Computing, 13, 67 

\bibitem[Tully et al.(2009)]{2009AJ....138..323T}
{Tully}, R.~B., {Rizzi}, L., {Shaya}, E.~J., {et~al.} 2009, \aj, 138, 323


\bibitem[Wither et al.(2011)]{Wither2011} Wither, J., Tsai, Y., Azuma, R. T. 2011, Computers and Graphics, 35, 4

\bibitem[Zhang \& Liu(2014)]{ZhangLiu2014} Zhang, F. and Liu, Feng. 2014, Proc. 2014 IEEE Conf. on Computer Vision and Pattern Recognition, 3262



\end{thebibliography}


\clearpage


\begin{deluxetable}{ll}
\tablecolumns{2}
\tablecaption{Rendering Engine Comparison\label{enginetable}}
\tablewidth{0pc}
\tabletypesize{\scriptsize}
\tablehead{ 
	\colhead{\textbf{Engine Name}} & 
	\colhead{\textbf{Support Notes}} 
	\\
}
\startdata
\textit{Render Default} &    All mesh material modes supported. \\
\textit{Cycles}         &    Halo materials for catalogs not supported.  Benefits include using node compositor.  \\
\textit{Game Engine}   &    Physics-based simulation and particle generator.  \\
\enddata
\tablecomments{The Blender Game Engine is mentioned here for completeness.  While it has many
important simulation properties that readers are encouraged to explore, it does not play a role
in the panoramic rendering at this time.
}
\end{deluxetable}

\clearpage

\end{document}